\documentclass[a4paper,11pt]{article}
\usepackage{color}
\usepackage{amsmath}    
\usepackage{amssymb}
\usepackage{graphicx}   
\usepackage{subfigure}
\usepackage{longtable}
\usepackage[utf8]{inputenc} %Codificacion utf-8
\usepackage[margin=3cm]{geometry}
\title{Ranking species in mutualistic networks}

\author{Virginia Dom{\'\i}nguez-Garc{\'\i}a$^{1}$ and Miguel A. Mu\~noz$^{1,\ast}$\\
  \normalsize{$^{1}$ Departamento de Electromagnetismo y F\'isica de la Materia,}\\
  \normalsize{and Instituto Carlos I de F\'isica Te\'orica y Computacional}\\
  \normalsize{ Universidad de Granada, 18071 Granada, Spain.}\\
  \small{$^{\ast}$Corresponding author: mamunoz@onsager.ugr.es}}

\begin{document}
\maketitle

\begin{abstract}
  Understanding the architectural subtleties of ecological networks,
  believed to confer them enhanced stability and robustness, is a
  subject of outmost relevance.  Mutualistic interactions have been
  profusely studied and their corresponding bipartite networks, such
  as plant-pollinator networks, have been reported to exhibit a
  characteristic ``nested'' structure.  Assessing the importance of
  any given species in mutualistic networks is a key task when
  evaluating extinction risks and possible cascade effects.  Inspired
  in a recently introduced algorithm --similar in spirit to Google's
  PageRank but with a built-in non-linearity-- here we propose a
  method which  --by exploiting their nested architecture-- allows us
  to derive a sound ranking of species importance in mutualistic
  networks. This method clearly outperforms other existing ranking
  schemes and can become very useful for ecosystem management and
  biodiversity preservation, where decisions on what aspects of
  ecosystems to explicitly protect need to be made.
\end{abstract}

\section*{Introduction}
Assessing the stability and robustness of complex ecosystems is a
fundamental problem in conservation ecology
\cite{Dunne2002,Dunne2004,Dunne2007,Sole2006,montoya}. The loss of an
individual ``keystone'' species can induce cascade effects --i.e.  a
series of secondary extinctions triggered by the primary one--
propagating the damage through the network. Thus, the relative
``importance'' of a given species within a ecological network could be
gauged as a function of the eventual size of the cascade of
extinctions its loss would potentially cause.  A successful ranking of
species importance should rank first those species that trigger larger
extinction cascades.

In the context of food webs, species rankings have been long sought
(see e.g.  \cite{Estrada,centrality}). For example, Allesina and
Pascual \cite{allesina} successfully applied the Google's PageRank
algorithm \cite{Page} to order species within food webs, much as Google
ranks webpages.

\begin{figure}[h]
\begin{center}
  \includegraphics[width=10.cm]{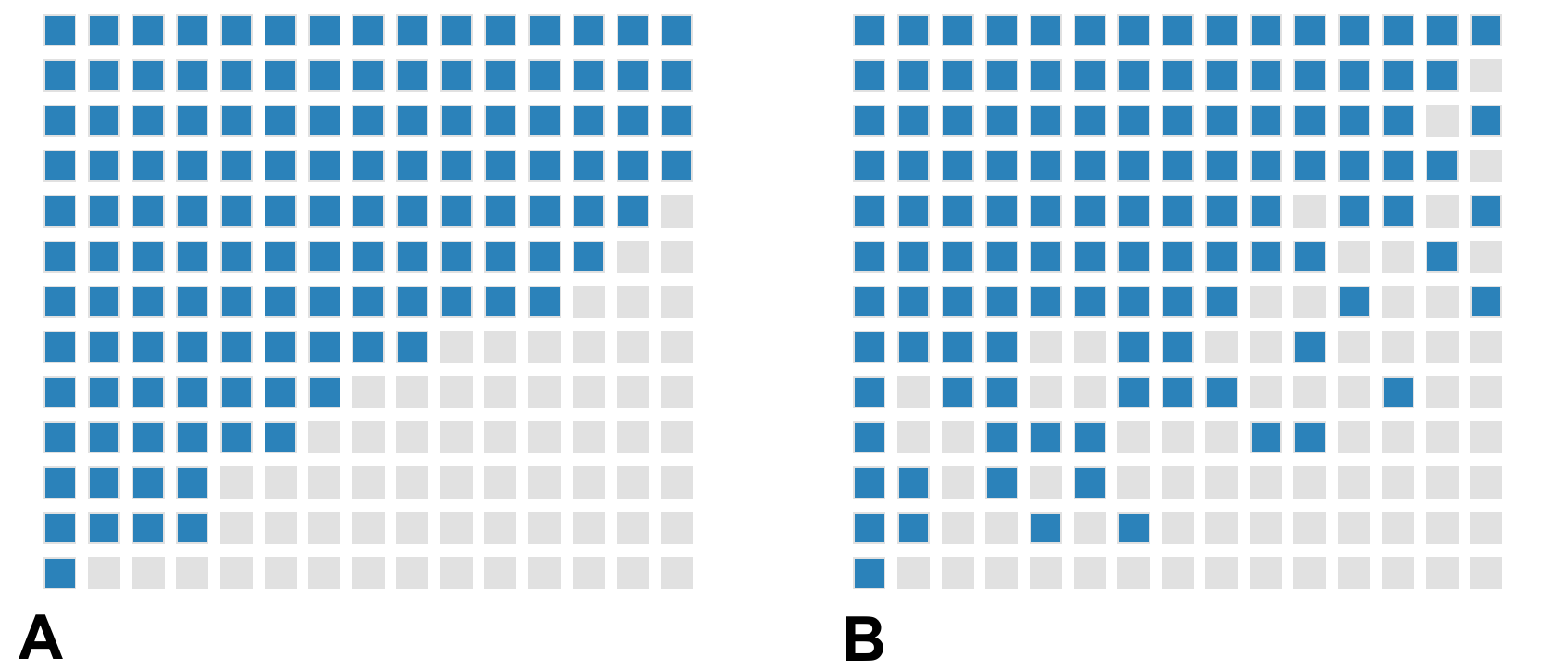}
  \caption{Example of two different bipartite networks with different
    levels of nestedness.  For simplicity, we focus on binary
    networks: blue squares correspond to existing interactions while
    empty ones describe absent links. A perfectly nested network (A)
    shows a characteristic interaction matrix in which
    \textit{specialist} species --with low connectivity-- interact
    only with \textit{generalist} ones. The matrix in (B) has a lesser
    degree of nestedness (see \cite{plos_nestedness,ghost} and
    \cite{calculator} for quantification of nestedness).}
\label{fig_nestedness_index}
\end{center}
\end{figure}

Mutualistic ecological communities such as those formed by plants and
their pollinators, plant seeds and their dispersers, or anemone and
the fishes that inhabit them, etc.  constitute another broadly studied
set of ecological networks. These comprise two different sets of
living beings that benefit from each other and as such can be
represented in terms of bipartite networks
\cite{Newman_rev}. Mutualistic networks turn out to have a very
particular ``nested'' architecture \cite{Basc-PNAS,Basc-Nature,ghost}
in which specialist species --interacting with only a few mutualistic
partners-- tend to be connected with generalists
(Figure \ref{fig_nestedness_index}).  Such a nested design is believed to
confer robustness against species loss and other systemic damages,
thus fostering biodiversity \cite{kaiser_bunbury_2,robust}.

Determining a ranking of species importance in mutualistic communities
poses an important practical challenge, as it would be highly
desirable to know which species are more crucial for the long-term
stability of the community. The goal would be to establish a proper
ordering of species, ranking them in order of decreasing importance
for the community. This would facilitate the design of sound
conservation policies protecting the most important species.

Following the experience from food webs we could, in principle, employ
the PageRank algorithm to rank mutualistic species in bipartite
networks.  PageRank \cite{Page,allesina} is a linear-algebra iterative
algorithm which, in a nutshell, computes the ``importance'' of a given
node as the linear superposition of the importance of the nodes
connecting to it, in a recursive and self-consistent way.  However, in
this work, taking inspiration from a recent breakthrough on
economics/econometrics \cite{pietronero1,pietronero2}, we propose to
employ a novel {\bf non-linear} algorithm specially designed for
bipartite networks.

Tacchella {\it et al.} \cite{pietronero1,pietronero2} analyzed
economic data from the world trade network (i.e. the bipartite network
of countries and the products they export). The goal was to infer an
objective ranking of countries in terms of their ``fitness'' and a
classification of the products in terms of increasing ``complexity''.
Inspection of such economic data reveals that rich (high fitness)
countries are not specialized into producing complex products (such as
high-tech devices) exclusively. Rather, they export a highly
diversified variety of goods, including less-complex ones
(e.g. cereals). On the other hand, poor (low fitness) countries only
produce low-complexity merchandises. These facts are reflected in the
nested structure of the corresponding bipartite network
\cite{pietronero1,pietronero2}, with a shape similar to that in Figure
\ref{fig_nestedness_index}. The main idea behind the novel algorithm
of Tacchella {\it et. al.}  is that while the fitness of a country can
be safely defined as the linear average of the complexity of the
products it exports, the reverse does not make sense. Indeed, the
complexity of a given product cannot be meaningfully estimated as the
average fitness of the countries producing it, but is much better
characterized by the minimal fitness required to produce it
\cite{pietronero1}. To implement this idea Tacchella {\it et al.}
proposed an iterative non-linear algorithm (see below) and were able
to compute the fitness of all countries and the complexity of all
products in a self-consistent way, using solely information contained
in the matrix of economic transactions. The novel algorithm clearly
outperforms PageRank and leads to striking implications for
understanding the global trade market \cite{pietronero1,pietronero2}.

Here, we consider a set of $63$ real mutualistic networks --all
  of them with a characteristic nested structure-- taken from the
literature ($45$ pollination networks, $16$ frugivore seed-dispersal,
and $2$ other networks; see Table 1) and rank the species accordingly
to different criteria (such as node-connectivity, betweenness
centrality, PageRank, etc.) including the novel non-linear
algorithm. Each of the employed criteria leads to a different ranking
of species.  We analyze the quality of any of these orderings by
monitoring how fast the network collapses if species are sequentially
removed in order of decreasing ranking.  The best ranking is the one
for which the network breaks down more rapidly. Our conclusion is that
the non-linear algorithm clearly outperforms all others, thus
providing us with an efficient and powerful scheme to gauge the
relative importance of species in mutualistic communities.

\section*{Results}

\subsection*{The non-linear ranking algorithm for mutualistic networks}

Inspired by the work of Tacchella {\it et. al.},
\cite{pietronero1,pietronero2} we propose a novel ranking algorithm
for mutualistic networks of ecological relevance. We shall refer to it
as mutualistic species rank ({\bf MusRank}). To establish a common
terminology for plant-pollinator, seed-disperser, and anemone-fish
networks, we refer to plants, seeds and anemones as ``passive'' (P)
elements, while pollinators, dispersers, and fishes are their
``active'' (A) partners; rather than fitness and complexity now we use
the terms \textit{importance} and \textit{vulnerability}, for the two
emerging species rankings, respectively.  It is natural to identify
products with \textit{passive} components and countries with
\textit{active} ones (but the opposite identification can also be
made; see below).  We assume that the importance of an active species,
is determined by the number of its mutualistic passive partners, each
one weighted with its own vulnerability: the more partners and the
more vulnerable they are, the more important an active element is.

On the other hand, the vulnerability of a passive element will be
bounded by the less important species it interacts with. The rationale
behind this is that, given that mutualistic networks are nested,
specialized species tend to interact with generalists. If a passive
element interacts only with generalists it is most certainly a
specialist and therefore highly vulnerable as it can disappear if a
few generalists go extinct.

The non-linear algorithm, encoding these ideas, is summarized in
eq.\ref{eq_algorithm}.  The importance of active elements,
$I_{A=1,..., A_{max}}$, and the vulnerability of passive ones,
$V_{P=1,...,P_{max}}$, are computed at iteration $n$ as a function
of their values in iteration $n-1$ using the interaction (or
adjacency) matrix $M_{AP}$ as the only input:
\begin{eqnarray}\label{eq_algorithm}
  {\tilde{I}}^{(n)}_A= \sum_{P=1}^{P_{max}} M_{AP}V_P^{(n-1)} &\longrightarrow &
  {I_{A}}^{(n)}=
  \frac{{\tilde{I}_{A}}^{(n)}} { {\langle {\tilde{I}_{A}}^{(n)}
      \rangle}_A }  \nonumber \\
  {\tilde{V}_P}^{(n)}=\frac{1}{ \displaystyle \sum_{A=1}^{A_{max}}{M_{AP}
      \dfrac{1}{{I_{A}}^{(n-1)}}} } & \longrightarrow& 
  {V_P}^{(n)} = \frac{{\tilde{V}_P}^{(n)}}{ {\langle {\tilde{V}_P}^{(n)} \rangle}_P}. 
\end{eqnarray}
Here, as in the work by Tacchella {\it et al.}, the adjacency
  matrix is considered to take binary values, but generalization to
  allow for real values is straightforward.  In a first step (left),
intermediate values of the importance and vulnerability are calculated
for each species: the first as the average of vulnerabilities of its
partners and the second as the inverse of the average of its partners
inverse importances \cite{pietronero1}.  In a second step (right),
both values are normalized to their mean values. In this way, starting
from arbitrary initial conditions (e.g. ${I_A}^{(0)}=1$ for all $ A$
and ${V_P}^{(0)}=1$, for all $P$) the two-step transformation above is
iterated until a fixed point is reached  (let us remark that we
  make no attempt here to prove that such a fixed point actually
  exists nor to investigate the conditions for the convergence of the
  method; see \cite{convergence}).  Such a fixed point --which does
not depend on initial conditions-- defines the output of the
algorithm: a ranking of importances and vulnerabilities for active and
passive species, respectively.

\subsection*{Assessing the quality of a given ranking}

\begin{figure}[h]
\begin{center}
\includegraphics[width=11.cm]{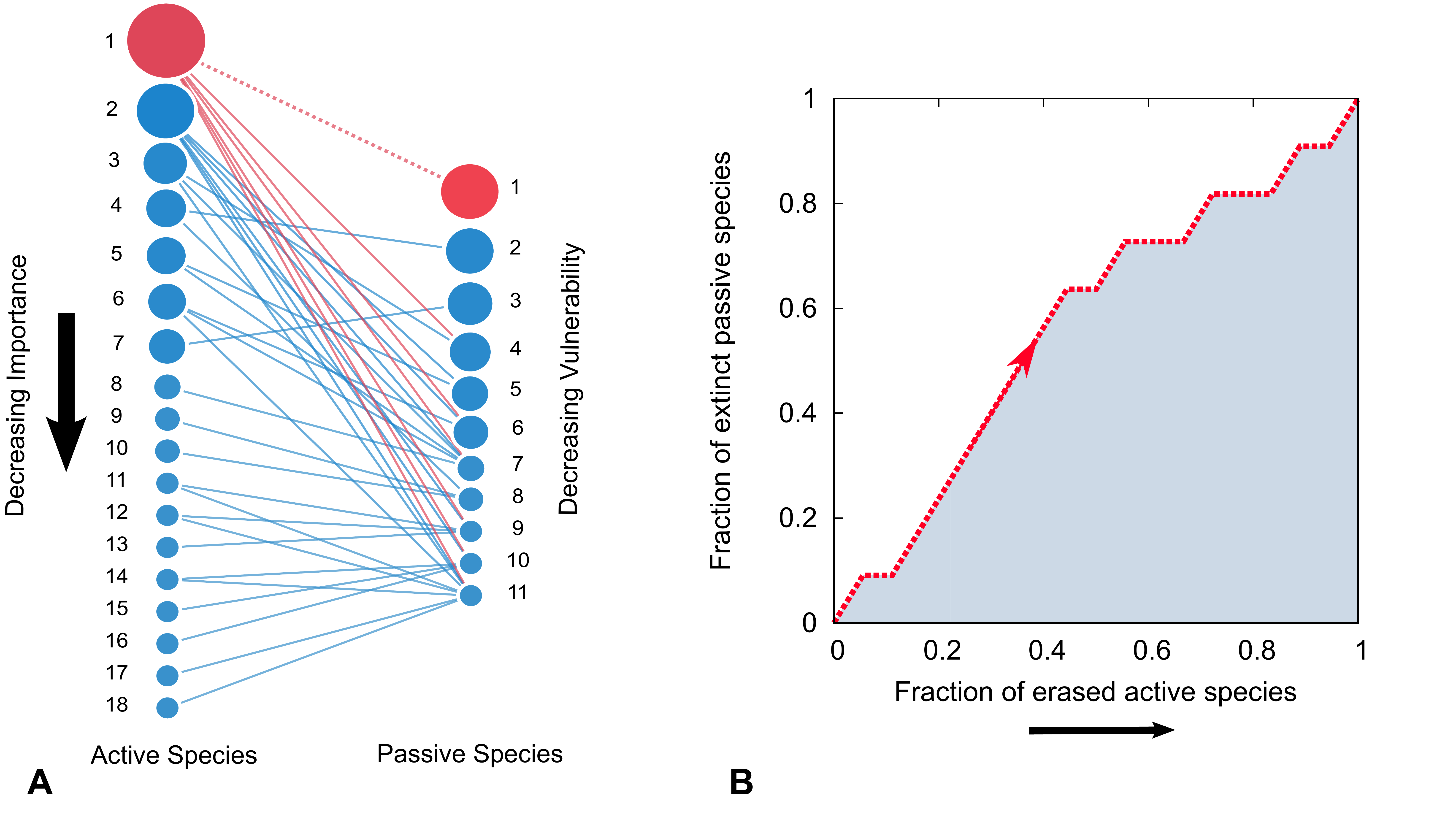}
\caption{Left: schematic representation of the extinction protocol for
  an empirical mutualistic network (Arctic community \cite{mos}) with
  $18$ active (pollinators) and $11$ passive (plants) species). Both
  active (left) and passive (right) species are ordered following some
  prescribed ranking; from the highest ranked species (top) to the
  lower-ranked ones (bottom). The (blue and red) lines represent
  mutualistic interactions as encoded in the interaction (or
  adjacency) matrix.  Active species are progressively removed from
  the community, their corresponding (red) links are erased, and
  passive species are declared extinct whenever they lose all their
  connections. Right: extinction curve, showing the fraction of
  extinct passive species as a function of the number of sequentially
  removed active ones for a given specified ranking. The shaded region
  is the extinction area for the ranking under
  consideration. Different rankings lead to different extinction
  areas. The larger the area the better the ranking.  }
\label{fig_extintions}
\end{center}
\end{figure}

In order to evaluate the quality of any possible ranking of species
for a given mutualistic network we proceed by computationally
implementing the following protocol (see Figure \ref{fig_extintions}
A).  Active species are removed progressively following the ordering
prescribed by any specified ranking algorithm. The ranking is
  kept fixed along this process, i.e. no re-evaluations of the ranking
  are performed once species are removed. Secondary extinctions are
monitored (a species is declared extinct when it no longer has any
mutualistic partners to interact with). The process is iterated until
all the species in the network have gone extinct. The total fraction
of extinct species as a function of the number of deleted species
defines a \textit{extinction curve} \cite{allesina}. For each possible
sequence of eradications the extinction area is obtained as the
integral of the extinction curve (see Figure \ref{fig_extintions} B).
This procedure allows for a quantitative discrimination of species
rankings: the best possible ordering of species would be the one for
which the largest extinction area is obtained upon progressively
removing active species in order of decreasing rank.

An exhaustive search of the optimal ranking (in the space of all
possible orderings) can be performed for relatively small networks but
becomes an unfeasible task for larger ones. To have an estimation of
the optimal ranking we implemented a genetic algorithm (\textbf{GA})
(see Methods) devised to obtain the maximal possible extinction area
by searching in the space of all possible orderings. For some of the
largest networks we studied (in particular, for Montane forest and
grassland, Beech forest and Phryganic ecosystem with $275, 678$, and
$666$ active nodes respectively; see table) the computational time
required for the genetic algorithm to converge is exceedingly large
and satisfactory results were not found.

Let us finally mention that we have also implemented a slightly
  modified version of the extinction protocol in which the ranking is
  re-evaluated after each species extinction. Beside being
  computationally much more expensive, this modified protocol leads, in
  general, to slightly worse results than the original one; however,
  even in this form MusRank outperforms all other rankings.

\subsection*{Algorithm testing and comparison with other rankings}

We compared different rankings based on (see Methods)
\ref{sec_algorithms}: a) decreasing closeness centrality
(\textbf{CLOS}), b) decreasing eigenvector centrality (\textbf{EIG}),
c) decreasing betweenness centrality (\textbf{BTW}), d) decreasing
degree centrality (\textbf{DEG}), e) increasing contribution to
nestedness (\textbf{NES}) as described in \cite{plos_nestedness}, f)
decreasing PageRank (\textbf{PAGE}), and g) decreasing importance as
measured by MusRank (\textbf{MUS}).

The average extinction area of the different algorithms was obtained
for all networks in the dataset. In the frequent case in which the
order is degenerate (more than one node were rated with the same
value), we considered $10^3$ different randomizations and computed the
averaged extinction area.

For the sake of completeness we have also repeated all the protocol
above, but exchanging in Eq.(\ref{eq_algorithm}) the roles of active
and passive species, i.e. assigning importances to passive species and
vulnerabilities to active ones. We refer to this as ``reversed''
algorithm.  We have also studied extinction areas by progressively
removing passive species (rather than active ones) and monitoring
secondary extinctions of active species.

\subsection*{Computational results}
\begin{figure}[tp]
\begin{center}
%\hspace*{-1cm}
\advance\leftskip-1cm
  \includegraphics[width=14cm]{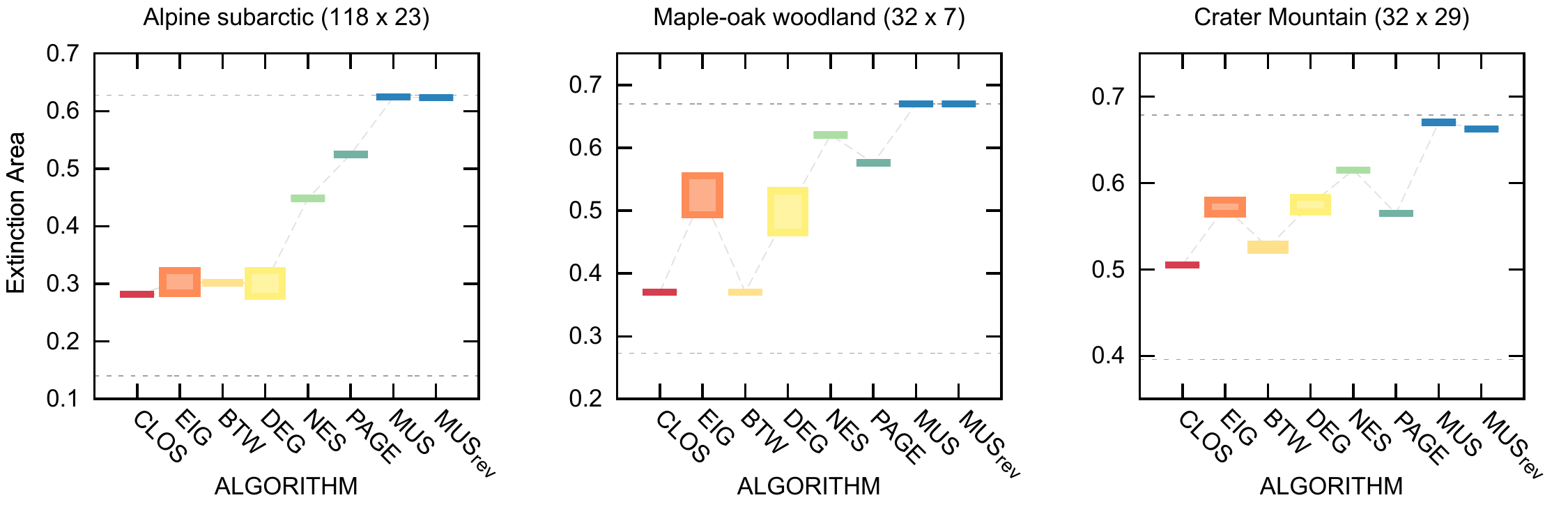}
  \caption{Extinction areas for three different mutualistic networks
    (names and sizes, specified above) as obtained employing the
    different ranking schemes described in the text. The upper dashed
    line shows the optimal performance corresponding to the ranking
    found by the genetic algorithm (\textbf{GA}) search, and the lower
    one the null-expectation, that is the averaged area obtained when
    targeting nodes in a random order.  The different algorithms used
    to rank the nodes are: closeness centrality (\textbf{CLOS}),
    eigenvector centrality (\textbf{EIG}), betweenness centrality
    (\textbf{BTW}), degree centrality (\textbf{DEG}), nestedness
    centrality (\textbf{NES}), PageRank (\textbf{PAGE}), and
    importance as measured by the MusRank
    (\textbf{MUS}).  \textbf{MUS$_{rev}$} corresponds to the
    reversed version of the algorithm in which the roles of active
    and passive species are exchanged.  The height of the boxes
    corresponds to the standard deviation of the results when
    averaging over $10^3$ random ways to break degeneracies in the
    orderings.}
\label{fig_area}
\end{center}
\end{figure}

Figure \ref{fig_area} illustrates the performance of the different
rankings/algorithms for three different instances of mutualistic
networks. Extinction areas are plotted for each of the considered
ranking algorithms. In the three cases MusRank gives results closest
to the corresponding optimal solutions as derived from the genetic
algorithm. In almost all of the $63$ studied cases, results are much
better for the novel ranking than for any of the other ones (see
Figure\ref{fig_area}).  PageRank gives similar results to MusRank in a
few cases (including a relative large network with $102$ nodes).
Apart from this, only for very small networks (with less than $17$
active species) some other method different from PageRank gives
extinction areas 

\begin{figure}[h!]
\begin{center}
\includegraphics[width=12.cm]{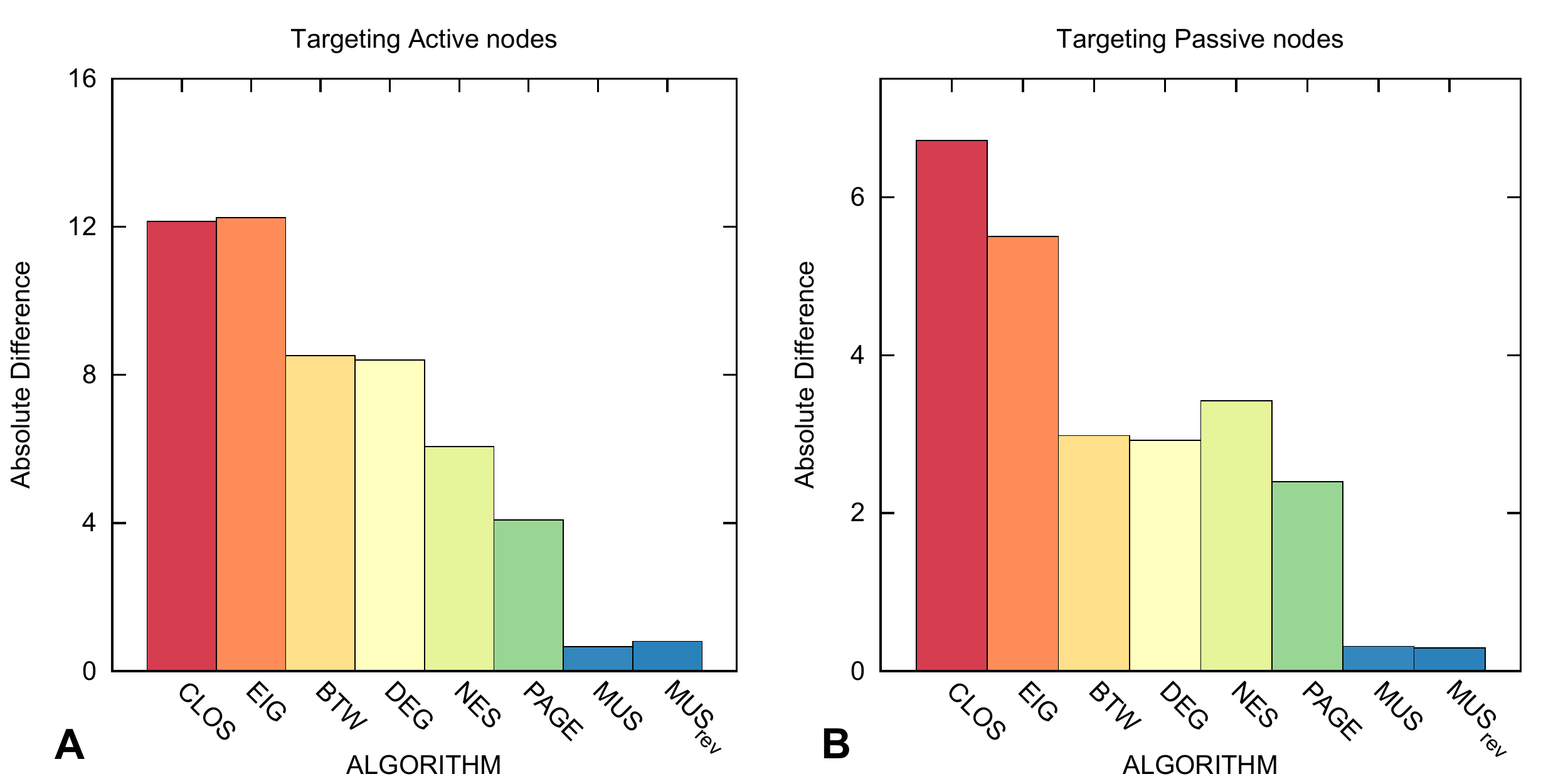}
\caption{Averaged deviation of the extinction area obtained for each
  of the employed rankings (or algorithms) from the maximal possible
  value as determined using the genetic algorithm (average over $60$
  networks in the database). The left A (right B) panel shows results
  when active (passive) species are targeted and passive (active)
  species undergo secondary extinctions. Results are consistently much
  better for the MusRank, in either the direct or the reversed
  version, than for any other ranking scheme.}
\label{fig_distances}
\end{center}
\end{figure}

\clearpage
similar to the ones of the novel algorithm. In about
one third of the networks, the ranking provided by MusRank is as good
as the one found by the \textbf{GA} and in some cases (networks for
which the \textbf{GA} could not converge in a reasonable time)
extinctions areas are larger for MusRank than for the \textbf{GA}.

Figure \ref{fig_distances} gives a global picture of the performance
of the different rankings. It shows the difference, averaged over $60$
mutualistic networks, between the optimal solution as found by the
\textbf{GA} and that of each specific ranking (the $3$ networks for
which the \textbf{GA} does not converge are excluded from this
analysis). Figure \ref{fig_distances}A illustrates that the ranking
provided by the MusRank --either in the direct or the reversed form--
greatly outperforms all others.

The same conclusion can be reached when progressively removing passive
rather than active species, ordered in a sequence of increasing
vulnerability (rather than decreasing importance), see Figure
\ref{fig_distances}B. Therefore, both targeting strategies and both
the direct and the reversed versions of the algorithm provide results
of similar quality.

\subsection*{Optimally packed matrices}

\begin{figure}[h]
\begin{center}\advance\leftskip-1cm
\includegraphics[width=14cm]{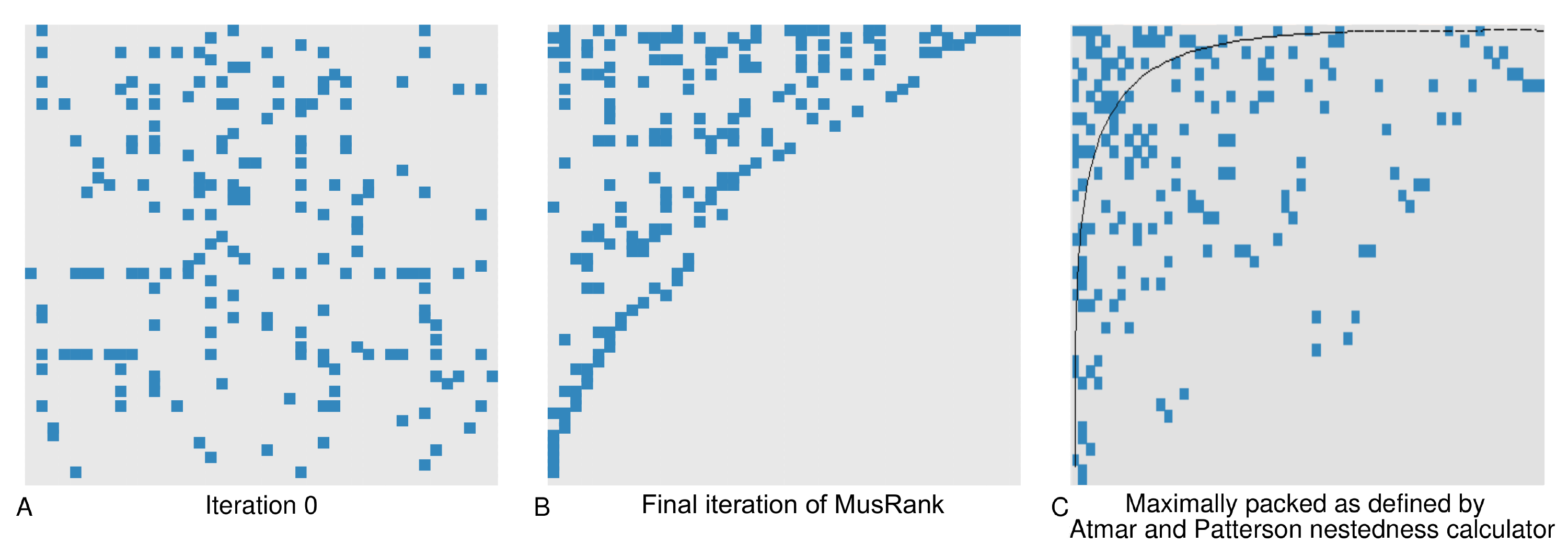}
\caption{Interaction matrix of a mutualistic community in the Andes
  \cite{arr_1} composed of $42$ pollinators and $61$ plants ordered by
  decreasing importance and increasing vulnerability respectively, as
  measured by MusRank. Panels A and B show two different shots of the
  iteration process: the initial random condition and the final
  (fixed-point) ranking obtained after iteration. Panel C shows the
  same matrix but with nodes labeled in an order which gives the
  maximally packed matrix according to the nestedness calculator of
  Atmar and Patterson \cite{calculator}. The novel algorithm provides
  a much more ``packed'' matrix than this frequently employed method.}
\label{fig_mapping}
\end{center}
\end{figure}

The ranking provided by MusRank, in which nodes are arranged by their
level of importance or vulnerability, permits us to obtain a highly
packed matrix as illustrated in Figure \ref{fig_mapping}. By
``packed'' we mean that a neat curve separates densely occupied and
empty parts of the matrix. It could be thought that this ordering
might be somewhat similar to the one that allegedly packs the matrix
in the most efficient way (as defined by existing algorithms usually
employed in the literature to measure nestedness
\cite{calculator}). However, as Figure \ref{fig_mapping} vividly
illustrates, the ordering provided by MusRank gives a more packed
matrix than that obtained by the standard method employed by
nestedness calculators \cite{calculator}. This suggests that MusRank
should be used (rather than existing ones) to measure nestedness in
bipartite matrices.

\section*{Discussion}
In this paper we have presented a novel framework to asses the
relative importance of species in mutualistic networks. Inspired by a
recent work on economics/econometrics we employ an algorithm, similar
in spirit to Google's PageRank but of non-linear character, that we
have named MusRank. The algorithm provides two complementary rankings:
one for active species (such as insects, birds, fish,...)  in terms of
their importance and one for passive species (plants and their seeds,
anemone, etc) in terms of their vulnerability. We also propose a
criterion to assess the quality of any given ranking of species: 
good rankings lead to a fast break-down of the corresponding mutualistic network 
when species are progressively removed in decreasing ranking order.

In most of the empirical mutualistic networks we have analyzed the use
of our novel framework rendered a ranking which clearly outperforms
all the alternative ones used as workbench.  Results are robust in the
sense that different implementations lead to similar rankings. In many
cases, the resulting ordering coincides or is very close to the
optimal one as found by a -computationally very costly- genetic
algorithm. Moreover, MusRank is much faster and finds excellent
rankings even for large mutualistic networks for which the genetic
algorithm is not able to find optimal solutions in a reasonable
computational time.  Therefore, the emerging ranking allows for
assessing the importance of individual species within the whole system
in a meaningful, efficient and robust way.  We conclude that rankings
of species importance in mutualistic networks should be constructed
employing MusRank.

Furthermore, as a by-product, the excellent packing of nested matrices
provided by this non-linear approach (see Figure \ref{fig_mapping})
calls for a redefinition of the way in which nestedness is
measured. In particular we suggest that nestedness calculators should
use the ranking provided by the present algorithm, which clearly
outperforms others in making the nested architecture evident.
Indeed, we believe that the nested structure of mutualistic
  networks is essential for the success of MusRank; it remains to be
  seen what is the performance of this scheme for bipartite networks
  without a nested architecture.

The novel approach --introduced here for the first time in the context
of mutualistic ecological networks-- may prove of practical use for
ecosystem management and biodiversity preservation, where decisions on
what aspects of ecosystems to explicitly protect need to be made.

\section*{Methods}
\subsection*{Algorithms}\label{sec_algorithms}
\begin{itemize}

\item \textbf{CLOS}: Nodes are sorted in order of decreasing closeness
  centrality. The closeness centrality of a node is measured as the
  inverse of the average shortest distance to all other nodes in the
  network. We computed it using the $closeness\_centrality$ function
  of the bipartite section of algorithms of the Python package
  \textit{NetworkX}.

 \item \textbf{EIG}: Nodes are sorted in decreasing order of their
  overlap with the highest eigenvalue. To calculate the eigenvector
  centrality of the bipartite network we used the $gsl$ functions for
  solving non-symmetric matrices.
  
 \item \textbf{BTW}: Nodes are sorted in order of decreasing betweenness
  centrality. The betweenness centrality of a node measures the
  fraction of shortest paths between all possible node pairs in the
  network, in which it appears.  We used the $betweeness\_centrality$
  function of the \textit{bipartite} section of algorithms in the
  Python package \textit{NetworkX}.
  
\item \textbf{DEG}: Nodes are sorted in order of decreasing number of
  connections. 

\item \textbf{NES}: Nodes are sorted in order of the inverse
  contribution to network nestedness. We calculate the total
  nestedness of a given bipartite matrix, and the contribution of each
  species to the total as described in \cite{plos_nestedness}. Species
  that contribute most to the community nestedness are the most
  vulnerable ones \cite{basc-contributors}. In order to look for the
  fastest community collapse we target them in order of increasing
  contribution to nestedness .
  
\item \textbf{PAGE}: Nodes are sorted in decreasing order of Google's
  PageRank.  The ranking is given by the projection over each node of
  the leading eigenvalue of the matrix $H$, whose elements are defined
  as $$h_{ij}=d \cdot a_{ij}/\sum_{j} a_{ij} + (1-d)/N.$$ The constant
  $d$ is a ``damping factor'' needed to warrant that the matrix is
  irreducible, and $a_{ij}$ are the elements of the adjacency matrix. 
  The value of $d$ has been set to $0.999$, but results
  are not very sensitive to this choice.

\end{itemize}

\subsection*{Genetic algorithm}

\textbf{GA}: The genetic algorithm is designed to seek for those
sequences of extinction that maximize the extinction area. We start
with $10^4$ different random orderings of the $A_{max}$ active species.
At each iteration-step two of these orderings are randomly selected.
Each one beats the other with a probability proportional to its
associated extinction area (normalized to the sum of both
extinction areas). The loser sequence is erased from the set and a
copy of the winner will occupy its place. With a small probability,
$\mu=0.005$, this copy suffers a mutation, meaning that two random
nodes exchange their positions in the ordering. The algorithm is
iterated until no better solutions are found in a sufficiently large
time window, that is, until no appreciable changes are seen in the
extinction area with increasing time. If the network is too large,
this algorithm might not be able to find a stationary optimal solution
within a reasonable computation time.

\bibliographystyle{naturemag}

%\end{thebibliography}
 \vspace{2cm}

\newpage

\section*{Acknowledgments:} 
We are grateful to Paolo Moretti and for a critical reading of the
manuscript. 

\section*{Author Contributions:}
Conceived and designed the work: VDG MAM.  Performed the experiments:
VDG.  Analyzed the data: VDG MAM.  Contributed
reagents/materials/analysis tools: VDG MAM.  Wrote the paper: VDG MAM.

 \vspace{4cm}

\section*{TABLES}
%%%%%%%%%%  TABLES %%%%%%%%%%%%%%%%
\begin{center}
\LTcapwidth=\textwidth
 \begin{longtable}{@{}llll}
\label{tabla}
   Network Name & $A_{max}$ & $P_{max}$ & Label \\
   \hline
   \hline
   Plant-Pollinator communities\\
   \hline
 Andean scrub (elevation 1), Cordon del Crepo (Chile)  \cite{arr_1} & 99 & 87 & 1 \\
 Andean scrub (elevation 2), Cordon del Crepo (Chile  \cite{arr_1} & 61 & 42 & 2 \\
 Andean scrub (elevation 3), Cordon del Crepo (Chile)  \cite{arr_1} & 28 & 41 & 3 \\
 Boreal forest (Canada)  \cite{barrett} & 102 & 12 & 4 \\
 Montane forest and grassland (U.S.A.)  \cite{cle} & 275 & 96 & 5 \\
 Grassland communities in Norfolk, Hickling (U.K.)  \cite{dicks} & 61 & 17 & 6 \\
 Grassland communities in Norfolk, Shelfanger (U.K.)  \cite{dicks} & 36 & 16 & 7 \\
 High-altitude desert, Canary Islands (Spain)  \cite{dupont} & 38 & 11 & 8 \\%%%
 Alpine subarctic community (Sweden)  \cite{elb} & 118 & 23 & 9 \\
 Mauritius Island (un-published) & 13 & 14 & 10 \\
 Mediterranean shrubland, Doñana (Spain)  \cite{herrera} & 179 & 26 & 11 \\
 Arctic community (Canada)  \cite{hocking} & 86 & 29 & 12 \\
 Snowy Mountains (Australia)  \cite{ino} & 91 & 42 & 13 \\
 Heathland -heavily invaded- (Mauritius Island)  \cite{kaiser_bunbury_2} & 135 & 73 & 14 \\
 Heathland -no invaded- (Mauritius Island)  \cite{kaiser_bunbury_2} & 100 & 58 & 15 \\
 Beech forest (Japan)  \cite{kato} & 678 & 89 & 16 \\%%%%%%
 Lake Hazen (Canada)  \cite{kev} & 110 & 27 & 17 \\
 Multiple Communities (Galápagos Islands)  \cite{mc_mullen} & 54 & 105 & 18 \\
 Woody riverine vegetation and xeric scrub (Argentina)  \cite{medan} & 72 & 23 & 19 \\
 Xeric scrub (Argentina)  \cite{medan} & 45 & 21 & 20 \\
 Meadow (U.K.)  \cite{mem} & 79 & 25 & 21 \\
 Arctic community (Canada)  \cite{mos} & 18 & 11 & 22 \\
 Deciduous forest (U.S.A.)  \cite{mot} & 44 & 13 & 23 \\
 Coastal forest, Azores Island (Portugal) \cite{ole_aigr} & 12 & 10 & 24 \\%%%%%%%
 Coastal forest, Mauritius Island (Mauritius)  \cite{ole_aigr} & 13 & 14 & 25 \\
 Coastal forest, Gomera Island (Spain)  (un-published) & 55 & 29 & 26 \\
 Upland grassland (South Africa)  \cite{ollerton} & 56 & 9 & 27 \\
 Coastal scrub (Jamaica)  \cite{percival} & 36 & 61 & 28 \\
 Phryganic ecosystem (Greece)  (un-published) & 666 & 131 & 29 \\
 Mountain, Arthur's Pass (New zealand)  \cite{primark} & 60 & 18 & 30 \\
 Mountain, Cass (New zealand)  \cite{primark} & 139 & 41 & 31 \\%%%%
 Mountain, Craigieburn (New zealand)  \cite{primark} & 118 & 49 & 32 \\%%%%%
 Palm swamp community (Venezuela)  \cite{ram} & 53 & 28 & 33 \\
 Caatinga  (N.E. Brazil)  \cite{santos} & 25 & 51 & 34 \\
 Maple-oak woodland (U.S.A.)  \cite{schemske} & 32 & 7 & 35 \\
 Peat bog (Canada)  \cite{small} & 34 & 13 & 36 \\
 Temperate rain forests, Chiloe (Chile)  \cite{smith_ram} & 33 & 7 & 37 \\
 Evergreen montane forest, Arroyo Goye (Argentina)   \cite{vazquez_1} & 29 & 10 & 38 \\
 Evergreen montane forest, Cerro Lopez (Argentina)   \cite{vazquez_1} & 33 & 9 & 39 \\%%%%%
  Evergreen montane forest, Llao Llao  (Argentina)   \cite{vazquez_1} & 29 & 10 & 40 \\
 Evergreen montane forest, Mascardi (c)  (Argentina)   \cite{vazquez_1} & 26 & 8 & 41 \\
 Evergreen montane forest, Mascardi (nc) (Argentina)  \cite{vazquez_1} & 35 & 8 & 42 \\
 Evergreen montane forest, Quetrihue (c) (Argentina)   \cite{vazquez_1} & 27 & 8 & 43 \\
 Evergreen montane forest, Quetrihue (nc) (Argentina)  \cite{vazquez_1} & 24 & 7 & 44 \\
 Evergreen montane forest, Safariland (Argentina)  \cite{vazquez_1} & 27 & 9 & 45 \\
   \hline
\hline
   Seed-Disperser communities\\
   \hline
 Eastern forest, New Jersey (USA)  \cite{baird} & 21 & 7 & 46 \\
 Forest (Papua New Guinea)  \cite{beehler} & 9 & 31 & 47 \\
 Forested landscape, Caguana (Puerto Rico)  \cite{carlo} & 16 & 25 & 48 \\
 Forested landscape, Cialitos (Puerto Rico)  \cite{carlo} & 20 & 34 & 49 \\
 Forested landscape, Cordillera (Puerto Rico)  \cite{carlo} & 13 & 25 & 50 \\
 Forested landscape, Frontón (Puerto Rico)  \cite{carlo} & 15 & 21 & 51 \\
 Tropical rainforest, Queensland (Australia)  \cite{crome} & 7 & 72 & 52 \\
 Coastal dune forest, Mtunzini (South Africa)  \cite{frost} & 10 & 16 & 53 \\
 Forest, Santa Genebra Reserve T1.(Brazil)  \cite{galetti} & 18 & 7 & 54 \\%%%%%%
 Forest, Santa Genebra Reserve T2.(Brazil)  \cite{galetti} & 29 & 35 & 55 \\
 Submontane rainforest (Central Philippine Islands)  \cite{hamman} & 19 & 36 & 56 \\
 Mediterranean shrubland, Hato Ratón (Spain)  \cite{jordano} & 17 & 16 & 57 \\
 Rainforest, Krau Game Reserve (Malaysia)  \cite{lambert} & 61 & 25 & 58 \\
 Crater Mountain Research Station (Papua New Guinea)  \cite{mack} & 32 & 29 & 59 \\
 Atlantic forest (SE. Brazil)  \cite{wes} & 110 & 207 & 60 \\
 Montane forest (Costa Rica)  \cite{wheelwright} & 40 & 170 & 61 \\
   \hline
\hline
   Other communities\\
   \hline
  Anemone-fish interactions in coral reefs  \cite{anemone_2} & 26 & 10 & 62 \\
  Ant-plant interaction in rainforest (Australia)  \cite{bluthgen} & 41
  & 51 & 63 \\
\hline
\hline
  \caption{Dataset of different mutualistic networks used throughout the study, with  $A_{max}$ active and $P_{max}$ passive species. } \label{network_data}
 \end{longtable}
\end{center}
%\end{landscape}

%%%%%%%%%%%%%%%%%%%%%%%%%%%%%%%%%%%%%%%%%%%%%%%%%%%%%%%%%%%%%%%%%%%%%
\end{document}